\documentclass[12pt]{iopart}
\usepackage{graphicx}
\begin{document}

\title[The truncated Holstein-Primakoff description of quantum rotors]{Virtues and limitations of the truncated Holstein-Primakoff description of quantum rotors}

\author{Jorge G. Hirsch,  Octavio Casta\~nos, Ram\'on L\'opez-Pe\~na, and Eduardo Nahmad-Achar}

\address{Instituto de Ciencias Nucleares,
Universidad Nacional Aut\'onoma de M\'exico \\
Apdo. Postal 70-543, Mexico D. F., C.P. 04510}
\ead{hirsch@nucleares.unam.mx}
\date{\today}
%\classification

\begin{abstract}
A Hamiltonian describing the collective behaviour of N interacting spins can be mapped to a bosonic one employing the Holstein-Primakoff realisation, at the expense of having an infinite series in powers of the boson creation and annihilation operators. Truncating this series up to quadratic terms allows for the obtention of analytic solutions through a Bogoliubov transformation, which becomes exact in the limit $N \rightarrow \infty$. The Hamiltonian exhibits a phase transition from single spin excitations to a collective mode. In a vicinity of this phase transition the truncated solutions predict the existence of singularities for finite number of spins, which have no counterpart in the exact diagonalization. Renormalisation allows to extract from these divergences the exact behaviour of relevant observables with the number of spins around the phase transition, and relate it with the class of universality to which the model belongs. 
In the present work a detailed analysis of these aspects is presented for the Lipkin model.
\end{abstract}



%\maketitle

\section{Introduction}

The dynamics of N two-level systems is well described by the Lipkin Hamiltonian \cite{lip65}. Born in nuclear physics, it has found extensive use in quantum optics, in the generation of squeezed states \cite{ueda}, multipartite entanglement~\cite{Vid04}, two-mode Bose-Einstein condensates \cite{zoller}, and monomolecular magnets \cite{Gat05}. It represents an approximation to ferromagnetic Ising models~\cite{jullien}, exhibiting a second-order phase transition in the limit of large number of particles which is well described by mean field techniques \cite{heiss2005,Cas05,Cas06}. In most cases a Holstein-Primakoff realisation is employed which, when truncated, provides analytical solutions in the thermodynamic limit \cite{vidal,Vid06}.  For a finite number of atoms, observables like the ground state energy, the energy gap and the number of excited atoms exhibit a singular behaviour at the phase transition, going to zero or to infinity, while numerical calculations show that they should remain finite. 

Similar singularities arise when using Holstein-Primakoff description of the rotor in more complex systems,  as for example in the Dicke model \cite{dicke}, where N two-level atoms are coupled with one electromagnetic mode  in a cavity \cite{Ema03,Dim07, nagy}.  Its energy eigenstates in the adiabatic limit, where the field frequency is much faster than the atomic excitation energy, are closely related with the collective spin model \cite{Lib10}.
While the limitations of truncated Hamiltonian and the need to include the next order corrections has been exposed previously \cite{vidal,Cas11a,Cas11b}, it seems necessary to discuss in detail the situations in which the prediction of the truncated Hamiltonian cannot be employed to describe finite systems, and also how useful information can be extracted from the %spurious 
divergences through the use of renormalisation techniques \cite{vidal}.

In this work the mean field description of the Lipkin model is obtained using the Holstein-Primakoff realisation of the quantum rotor. It is shown that coherent Heisenberg-Weyl states \cite{Glau} provide a mean field description of the system, exhibit the existence of different phases in the parameter space, associated with different qualitative descriptions, and phases transitions between them. Analytical expressions are obtained for the ground state energy and for the expectation values of the number of photons and of excited atoms, which are intensive and are shown to represent the thermodynamic limit. A truncated Hamiltonian is built which has analytic solutions, and provides the eigenstates of the system far from the phase transitions. Close to them, the observables exhibit the singular behaviour mentioned above. The spurious character of these singularities, and the way in which relevant critical exponents can be extracted from them are discussed in detail. The general aim it to present the above mentioned subjects in a pedagogical form. We follow closely the treatment of Holstein-Primakoff, used in a different but equivalent way by Dusuel and Vidal \cite{vidal}, and the analysis of the fidelity susceptibility presented by Gu \cite{gu10} and Zanardi \cite{Zan07}, with innovations in both the expansion and the obtention of the critical exponent employing the fidelity susceptibility.

\section{The Lipkin Hamiltonian}

 The Lipkin Hamiltonian \cite{lip65} describes the collective behaviour of $N$ spins or two-level atoms, with energy separation $\epsilon$, which interact by scattering pairs of particles between the two levels. In the quasi-spin formalism it has the form
	\begin{equation}
		H= \epsilon  J_z +\frac{ \gamma _x }{N} J_x^2 + \frac{ \gamma _y }{N} J_y^2 ,
		\label{Ham}
	\end{equation}
where $J_x, J_y, J_z$ are the three components of the angular momentum operator, with the usual commutation relations $[J_j,J_k] = i \varepsilon_{jkl} J_l$, and $\gamma_x, \gamma_y$ are the coupling strenghts.

\subsection{Mean Field description}

In the literature it is customary to employ the Holstein-Primakoff representation of the angular momentum operators \cite{Hol40}
\begin{equation}
J_+ =\sqrt{N}\, b^{\dagger }\sqrt{1- \frac{b^{\dagger } b}{N}}, \,\,
J_- =\sqrt{N} \sqrt{1- \frac{b^{\dagger } b}{N}}\, b, \,\,
J_z = \, b^{\dagger}  b - \frac N 2, 
\label{hpmap}
\end{equation}
where $J_+ = J_x + i J_y, \, J_- = J_x - i J_y$, the Bose operators  $b^{\dagger},  b$ obey the commutation relation $[b, b^{\dagger}] = 1,$ and the vacuum $|\, \rangle$ of the bosons satisfies $b |\, \rangle = 0$.
Making these substitutions into $H$, Eq. (\ref{Ham}), the bosonic Hamiltonian is built. Care must be taken in using the commutation relations to move the creation operators (outside the square roots) to the right, and the annihilation ones to the left. 

The mean field  description of this Hamiltonian is easily obtained employing as a trial state the
Heinsenberg-Weyl coherent state $\vert \alpha \rangle$, which is the eigenstate of the bosonic annihilation operator
$
b\, \vert \alpha \rangle = \alpha \, \vert \alpha \rangle ,
$
where $\alpha = \rho\, e^{i \phi}$ is a complex numbers, with $\rho \geq 0$ and $0 \leq \phi < 2 \pi$.

The expectation value of the Hamiltonian for this coherent state provides the {\em energy surface}. Employing the approximation 
$\sqrt{1- \frac{b^{\dagger } b}{N}}|\beta \rangle \approx \sqrt{1- \frac{\rho ^2}{N}}|\beta \rangle $ \cite{hir12}, 
which becomes exact in the thermodynamic limit, when the number of atoms goes to infinity, it takes the simple form \cite{Kap91}
\begin{eqnarray}
\langle \alpha | H |\alpha \rangle =&\epsilon  \left(\rho ^2 \,\,- \frac{N}{2}\right)+\frac{ \gamma _x+\gamma _y }{4} \left(1- \frac{\rho ^2}{N}\right)\left(1+2\rho ^2\right)\\
&+\frac{ \gamma _x-\gamma _y }{2} \left(1- \frac{\rho ^2}{N}\right)\rho ^2\,\,\cos[2\phi ] \nonumber
\end{eqnarray}		
For a given set of Hamiltonian parameters  $ \epsilon,\gamma_x,\gamma_y$, the values $ \rho_{c}, \phi_{c}$ which minimise this expression provide the mean field wave function.
%, the best approximation to the exact ground state as a product wave function. 
They are obtained by solving the equations for the energy surface critical points
\begin{eqnarray}
\frac {\partial \langle \alpha \vert H \vert \alpha \rangle}{\partial \rho} = 0, ~
 \frac {\partial \langle \alpha  \vert H \vert \alpha  \rangle}{\partial \phi} = 0.
 \end{eqnarray}
The solutions of these equations associated with the minima of the energy surface are \cite{Cas05,Cas06}
\begin{eqnarray}
\left\{\begin{array}{ccc}
 \rho_{c} = 0,  &\phi_{c} \mbox{\small ~~undetermined,}  &\mbox{ if }  \gamma_x  \ge \gamma_c  \mbox{ and }  \gamma_y  \ge \gamma_c,  \\
 \rho_{c} =  \, \sqrt{\frac{N}{2}\left(1- \frac{\gamma_c }{\gamma _x}\right)}, &\phi_{c} = 0, \pi ,&\mbox{ if }   \gamma_x  < \gamma_c \mbox{ and }  \gamma_x  \le \gamma_y,\\
  \rho_{c} =  \, \sqrt{\frac{N}{2}\left(1-\frac{\gamma_c }{\gamma _y}\right)}, &\phi_{c} = 0, \pi ,&\mbox{ if }   \gamma_y  < \gamma_c \mbox{ and  }  \gamma_x  > \gamma_y,
 \end{array}
 \right.
 \label{rhoc}
\end{eqnarray}
with $\gamma_c \equiv -\epsilon$.
The first case, $\gamma_x  \ge \gamma_c $ and $\gamma_y  \ge \gamma_c$, defines the {\it normal} region (I), where in the ground state all atoms are in their lowest energy state. In the first {\em deformed} region (II), where $\gamma_x  < \gamma_c$ and $\gamma_x \le \gamma_y$, it is energetically favoured to collectively excite all atoms, and the ground state is doubly degenerate, as there are two critical phases  $\phi_{c} = 0, \pi$. The second deformed region (III), where $\gamma_y  < \gamma_c$ and $\gamma_x > \gamma_y$, is symmetric to the first one:  it is obtained by interchanging $J_x \leftrightarrow J_y$. In what follows we will restrict our analysis to regions I and II.  
   
Employing these critical values, explicit analytical expressions can be found for intensive observables: the ground state energy per atom $\varepsilon_{gs} = \langle \alpha_c \vert H \vert \alpha_c \rangle /N $, and the fraction of excited atoms  $n_e = 2 \rho_c^2 /N $:
\begin{eqnarray}
\varepsilon_{gs} =  \frac{\gamma_c}{2} + \frac{\gamma_x + \gamma_y}{4 N},  &n_e =0, &\mbox{\quad     region I }     \nonumber \\
\varepsilon_{gs} = \frac{\gamma_c^2 + \gamma_x^2 }{4 \gamma_x} 
+ \frac{ (\gamma_x + \gamma_c)( \gamma_y + \gamma_x )}{8 N \gamma_x},~~~
&n_e =1-\frac{\gamma_c }{\gamma_x},  &\mbox{ \quad  region II}.
   \label{egs}
\end{eqnarray}

\begin{figure}[h]
\scalebox{0.8}{\includegraphics{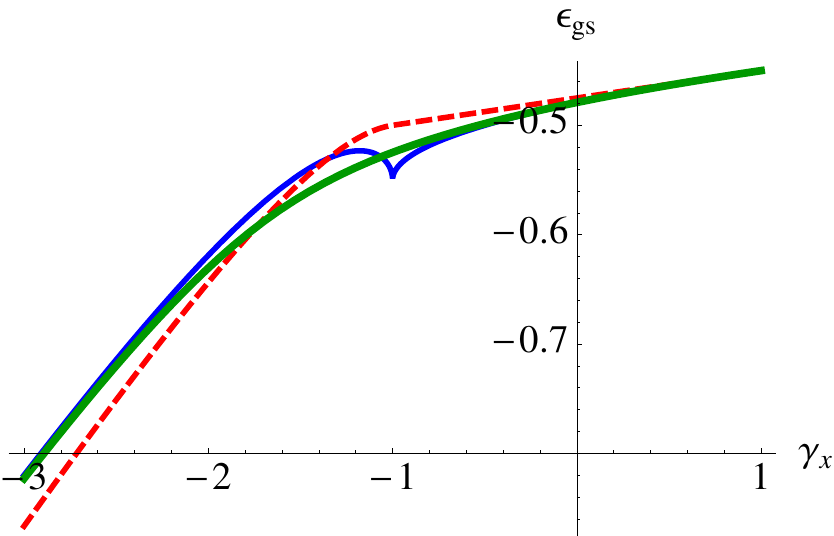}}\qquad\qquad
\scalebox{0.8}{\includegraphics{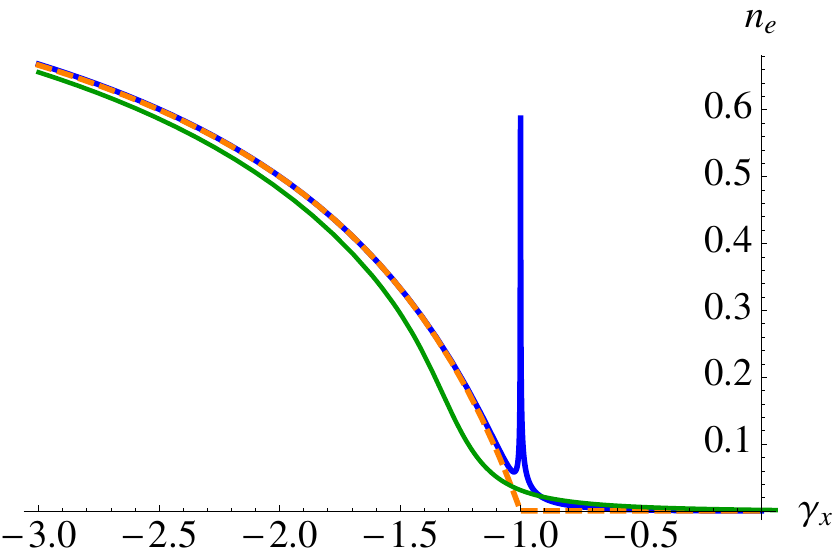}}\qquad\qquad
\caption{\label{energs}
(Color online) Left: Ground state energy per atom,  calculated through numerical diagonalization of the matrix Hamiltonian $H$ for N=  10 (continuous green line),  and from Eq. (\ref{egs}) (dashed red line). Right: Fraction of excited atoms calculated through numerical diagonalization of the matrix Hamiltonian  $H$ for N=  40 (continuous green line) and from Eq. (\ref{egs}) (dashed red line). Both plots were calculated using $\gamma_c = -1$ and $\gamma_y=1$. Peaked blue lines show the spurious results obtained with the truncated Hamiltonian (see next section). }
\end{figure}

Although they were obtained using different techniques, these mean field expressions, shown as dashed red lines in  Fig. \ref{energs}, exactly coincide with those presented in \cite{Vid04,Cas05}. They reproduce reasonably well  the numerical results even for a relatively small number of particles, as can be seen in Fig. \ref{energs}, where the numerical diagonalization of the Hamiltonian matrix $H$ is shown as a continuous green line, for N= 10 and 40 atoms, for $\gamma_c = -1$ and $\gamma_y=1$.
At this point it is worth to emphasize that the mean field description becomes exact in the thermodynamic limit, where only the first terms in the energy per atom remain finite. The peak in the continuous blue lines at $\gamma_x=-1$, which goes to infinity for the fraction of excited atoms, is similar to the one shown in Fig.1 of Ref. \cite{nagy}. 
%It is a spurious result coming from the truncated version of the Hamitonian, as explained in the next section.
It is an exact result coming from the truncated version of the Hamiltonian which has no counterpart in the numerical diagonalization of the Lipkin Hamiltonian (\ref{Ham}) for any finite number of atoms, as explained in the next section.

\section{Beyond mean field}

The first step in going beyond the mean field description is the introduction of the displaced boson operators, as in \cite{Ema03,Vid04},
\begin{equation}
c^\dagger = b^\dagger  - \rho_c,  \, \,c = b - \rho_c   , \label{disp}
\end{equation}
which are defined to satisfy the property  $c \vert \alpha_c \rangle = 0$. While in the normal phase, where $\rho_c=0$,  there is not displacement, using $c^{\dagger}, c$ allows for a general treatment of the Lipkin model beyond mean field. 

If the ground state is well described by the coherent state (the vacuum, in the case of the normal phase), it is valid to make the approximation
\begin{equation}
 \frac{ c^{\dagger}  c}{N}  \rightarrow  \frac{\langle c^{\dagger}  c \rangle}{N} \ll 1.
\end{equation}
%It must be stressed that this approximation fails badly close to the phase transition, where the truncated Hamiltonian cannot be employed. 
It must be stressed that this expansion becomes exact only in the thermodynamic limit, where $N \rightarrow \infty$. For any finite N it will have problems,  which become particularly relevant close to the phase transition.

Far from the phase transition region it is possible to expand the square roots in Eq. (\ref{hpmap}) in powers of $N$ series, conserving terms of order $N^1, N^{\frac 1 2}$ and $N^0$, and neglecting all negative powers of $N$. Care must be taken because $\rho_c$ is of order $N$. The truncated version $H^{(t)}$ of the Hamiltonian $H$, which is quadratic in the new bosons, is 
\begin{equation}
H^{(t)} = A + B \,c^{\dagger }c + C \left(c^{\dagger \,2}+c^2\right)
\label{ht}
\end{equation}
with
\begin{eqnarray}
A = -\gamma_c \left(\rho_c^2-\frac{N}{2}\right)+\frac{\gamma _x}{4 N}\left(N-3 \rho_c^2+4 \rho_c^2 \left(N-\rho_c^2\right)\right)+\frac{ \gamma _y}{4 N}\left(N-\rho_c^2\right) , \nonumber \\
B=-\gamma_c +\frac{\left(N-7 \rho_c^2\right) \gamma _x}{2 N}+\frac{\left(N-\rho_c^2\right) \gamma _y}{2 N}, \\
C =\frac{\left(N-5 \rho_c^2\right) \gamma _x}{4 N}-\frac{\left(N-\rho_c^2\right) \gamma _y}{4 N}. \nonumber
\end{eqnarray}
There is an extra term proportional to $c^{\dagger}+c$ whose coefficient vanishes exactly when employing the values of $\rho_c$ given in Eq. (\ref{rhoc}). The Hamiltonian $H^{(t)}$ can be diagonalized through the Bogoliubov transformation \cite{vidal}
\begin{equation}
c^{\dagger }= \cosh\left[\frac{\Theta }{2}\right] \, a^{\dagger } + \sinh\left[\frac{\Theta }{2}\right] \,a , \,\, c= \cosh\left[\frac{\Theta }{2}\right] \, a+ \sinh\left[\frac{\Theta }{2}\right] \, a^{\dagger},
\end{equation}
in terms of the new bosons $a^{\dagger}, a$. When replaced in Eq. (\ref{ht}),  it reads
\begin{eqnarray}
H^{(t)} = &A + B \sinh \left[\frac{\Theta }{2}\right]^2+ C \sinh [\Theta ]+(B \cosh[\Theta ]+ 2 C \sinh[\Theta ]) a^{\dagger }a  \nonumber\\
&+\left( \frac{B}{2} \sinh[\Theta ] +C \cosh[\Theta ]\right)\left(a^{\dagger \,2 }+ a^2\right).
\label{hta}
\end{eqnarray}
The last term cancels out by  selecting $ \tanh[\Theta ]=- 2\frac{C}{B} $. The truncated Hamitonian has the final diagonal form \cite{vidal}
\begin{eqnarray}
H^{(t)} = A+\frac{1}{2}\left(\sqrt{B^2- 4\, C^2}-B\right)+\sqrt{B^2- 4\, C^2} \, a^{\dagger }a = N \varepsilon_{gs}^{(t)} + \Delta \, a^{\dagger }a .
\label{hta2}
\end{eqnarray}
The ground state of the truncated Hamiltonian is the new vacuum $|0\rangle$, which satisfies $a |0\rangle=0$.
The coefficient of the last term is the gap $\Delta$, the energy separation between the ground and the first excited state in the normal phase, and between the ground and the second excited state in the deformed region, where the first excited state becomes degenerate with the ground state in the thermodynamic limit \cite{vidal,Ema03}. Its explicit form is
\begin{equation}
\Delta \equiv \sqrt{B^2- 4\, C^2} = \left\{ \begin{array} {cc} \sqrt{(\gamma_x-\gamma_c) (-\gamma_c +\gamma_y)} &\mbox{\quad     region I } , \vspace{.2cm}\\ 
\sqrt{\left(\gamma_x ^2-\gamma_c^2\right) (\gamma_x-\gamma_y) / \gamma_x}
 &\mbox{\quad     region II } . \end{array} \right.
 \label{gap}
\end{equation}
\begin{figure}[h]
\begin{center}
\scalebox{1.0}{\includegraphics{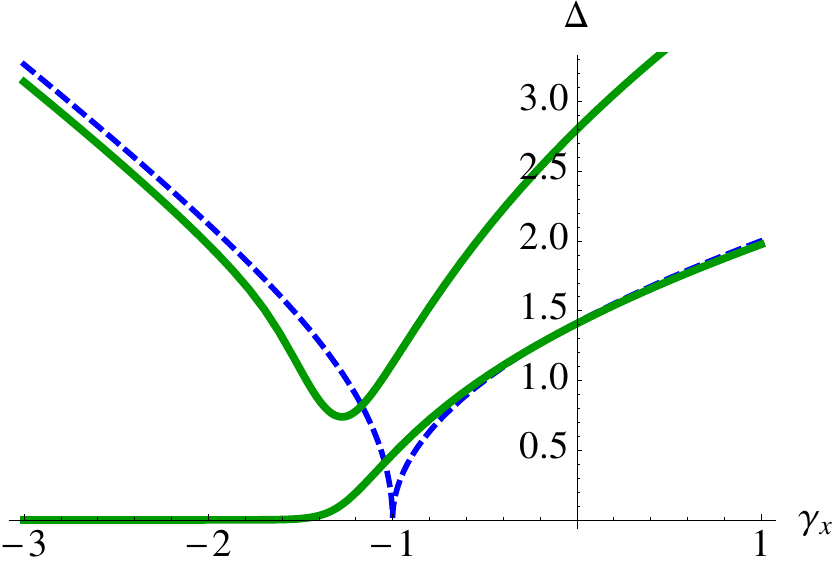}}
\end{center}
\caption{\label{fgap}
(Color online) The gap, Eq. (\ref{gap}), as function of $\gamma_x$, shown as a dashed blue line for $\gamma_c=-1$ and $\gamma_y=1$, and the first and second excitation energies, obtained numerically for N= 40, displayed as continuos green lines. }
\end{figure}

Figure \ref{fgap} plots the gap as a function of $\gamma_x$ as dashed blue line. It becomes null at the phase transition $\gamma_x = \gamma_c$. The excitation energies of the first and second excited states, obtained through exact diagonalization for N=40, are also displayed. The truncated Hamiltonian allows a good description of the gap, which becomes exact when $N\rightarrow \infty$. The minimum of the excitation energy is a precursor of the phase transition at finite N, which takes place at a different value of $\gamma_x$ for each $N$ \cite{Cas_jerry}. 

The ground state energy per atom $\varepsilon_{gs}^{(t)}$ is given by the constant terms in $H^{(t)}$, Eq.~(\ref{hta2}), divided by N
\begin{equation}
\varepsilon_{gs}^{(t)}  = \left\{ \begin{array} {cc} 
\frac{\gamma_c}{2} + \frac{\gamma_c}{2 N} + \frac {1}{ 2 N} \sqrt{( \gamma_x-\gamma_c) (\gamma_y-\gamma_c)}
&\mbox{\quad     region I } , \vspace{.2cm} \\
 \frac{\gamma_c^2 + \gamma_x^2 }{4 \gamma_x}+\frac{\gamma_x}{2 N} +  \frac {1}{ 2 N} \sqrt{\left(\gamma_x^2-\gamma_c ^2\right) (\gamma_x-\gamma_y) / \gamma_x}
 &\mbox{\quad     region II } . \end{array} \right.
 \label{egst}
\end{equation}
It is plotted as a continuous blue line in Fig. \ref{energs} (left), and displays a spike at $\gamma_x=\gamma_c$ which is %spurious, 
an artifact of the truncation, and vanishes as $N\rightarrow \infty$. 

The fraction of excited atoms is obtained by expressing $J_z$ in terms of the new bosons ($a^{\dagger},a$), and evaluating $n_e=\frac {2 \langle J_z \rangle}{N} + 1$ in the new vacuum. 
\begin{equation}
n_e^{(t)}  = \left\{ \begin{array} {cc} 
\frac{1}{2 N}  \frac{\gamma_x+\gamma_y-2 \gamma_c}{\sqrt{(\gamma_x -\gamma_c) (\gamma_y-\gamma_c)}} -\frac{1}{N}
&\mbox{\quad     region I } , \vspace{.2cm}\\
 \frac{\gamma_x -\gamma_c}{ \gamma_x }+\frac{ 1}{ 4\,N \sqrt{\gamma_x -\gamma_c}}\frac{\gamma_c \gamma_y+ \gamma_x (3 \gamma_c -5 \gamma_x +\gamma_y)}{ \sqrt{\gamma_x (\gamma_c +\gamma_x) (\gamma_x-\gamma_y)}} -\frac{1}{N} 
 &\mbox{\quad     region II } . \end{array} \right.
 \label{net}
\end{equation}
The curve $n_e^{(t)}$ vs. $\gamma_x$ is displayed as a continuous blue line in Fig. \ref{energs} (right). It follows closely the mean field prediction, except in a vicinity around $\gamma_x=\gamma_c$, where it diverges. In this case the pathologic behaviour survives in the thermodynamic limit, where the mean field expression becomes exact. It reminds us that in this region the solutions of the truncated Hamiltonian are no longer valid. There are, however, reliable ways to extract useful information from the spurious results obtained around the phase transition. They are the subject of the next section.

\section{Renormalisation and critical exponents}

The renormalisation procedure postulates that it is possible to extract the correct functional form $O(\gamma,N)$, smooth and finite for any finite N, of any observable $O^{(t)}(\gamma,N)$ which becomes singular at the phase transition \cite{cardy} when described  using the truncated Hamiltonian. If $O^{(t)}(\gamma,N)$ is singular as $ N^{\beta}(\gamma-\gamma_c)^{-\alpha}$,  employing a scaling function $\left[ N^{\nu}(\gamma-\gamma_c)\right]^\alpha $ the regular function is built as \cite{vidal}
\begin{eqnarray}
O(\gamma,N) &
= O^{(t)}(\gamma,N) \left[ N^{\nu}(\gamma-\gamma_c)\right]^{\alpha}\nonumber\\
& \rightarrow N^{\beta}(\gamma-\gamma_c)^{-\alpha} \left[ N^{\nu}(\gamma-\gamma_c)\right]^{\alpha} = N^{\beta + \alpha \, \nu},
\end{eqnarray}
where the power $\alpha$ in the second term was selected to cancel out the singularity at $\gamma_c$, and $\nu$ defines the {\em class of universality} to which the model belongs. In the Lipkin model numerical analysis points to $\nu = \frac 2 3$~\cite{jullien,Cas_jerry,gu10}. We will show that this number can be deduced analytically using the fidelity susceptibility. Also we will exhibit the singular behaviour of the point $(\gamma_x,\gamma_y)=(\gamma_c,\gamma_c)$ in the parameter region, where $\nu$ has a different value when approached along the lines $\gamma_y=\gamma_c$ or $\gamma_x=\gamma_c$.

The singular term in the energy per atom, which gives rise to the spurious spike when $\gamma_x \rightarrow \gamma_c$, behaves like $\sqrt{\gamma_x -\gamma_c}/N$, as shown in Eq. (\ref{egst}). Renormalizing it we obtain
\begin{equation}
\varepsilon_{gs}^{ren}(\gamma_x,N) \rightarrow  (\gamma_x -\gamma_c)^{\frac 1 2} N^{-1}  \left[ N^{\frac 2 3}(\gamma_x -\gamma_c)\right]^{- \frac 1 2} = N^{- \frac 4 3}.
\end{equation}
The singular term can be calculated numerically, substracting from the exact energy per atom the regular terms in Eq. (\ref{egst}). It goes to zero as function of N as predicted \cite{vidal}.  

The divergence in the fraction of excited atoms can be manipulated in the same way. Taking the divergent term from Eq. (\ref{net}), it gives
\begin{equation}
n_e^{ren}(\gamma_x,N) \rightarrow  (\gamma_x -\gamma_c)^{-\frac 1 2} N^{-1}  \left[ N^{\frac 2 3}(\gamma_x-\gamma_c)\right]^{\frac 1 2} = N^{- \frac 2 3}.
\end{equation}
In Ref. \cite{vidal} this dependence on N was also confirmed numerically.

\subsection{Fidelity susceptibility}

In classical information theory the fidelity measures the accuracy of a transmission \cite{joz94}. It also provides a powerful tool to study quantum phase transitions \cite{Zan07,gu10}. For a pure state $|\psi (\lambda)\rangle$ which varies as a function of a control parameter $\lambda$, the fidelity is defined as
\begin{equation}
F(\lambda, \lambda+\delta\lambda) \equiv \left| \langle \psi (\lambda) |\psi (\lambda+\delta\lambda)\rangle\right|^2 .
\end{equation}   
It diminishes when there are sudden changes in the wave function, which for finite N represent a precursor of a phase transition, and has a minimum where these changes are the largest. Even more sensitive is the fidelity susceptibility \cite{you07}, closely related to the second derivative of the fidelity. It has a maximum reflecting the phase transition. 
Expressing the Hamiltonian under study as $H = H_0 +\lambda H_I $, which makes explicit the dependence on the control parameter, the fidelity susceptibility can be calculated as \cite{gu10,Cas_jerry}
\begin{equation}
\chi_F = \sum\limits_{k\ne 0} \frac{\left| \langle \psi_k (\lambda)| H_I |\psi_0 (\lambda)\rangle\right|^2}{\left(E_k(\lambda) - E_0(\lambda)\right)^2},
\label{chif}
\end{equation}
 where $|\psi_k (\lambda) \rangle$ denotes the $k$ eigenstate of $H$ with energy $E_k(\lambda)$, and $k=0$ refers to the ground state. 
 
In our case, we select $H_I = J_x^2/N$ and study the dependence on $\gamma_x$. 
The way to proceed is to express it in terms of the bosons $a^{\dagger}, a$, as
\begin{equation}
J_x^2 = j_1 +j_2 a^{\dagger }a+j_3 \left(a^{\dagger }+a\right)+ j_4 \left(a^{\dagger \, 2} +a^2\right).
\end{equation}
The first two terms do not contribute to $\chi_F$. The third one connects the ground state with a one boson state, with energy $E_1 = \Delta$, the last one with a two boson state with energy  $E_2 =  2 \Delta$. 
In the normal phase, region I, we find
\begin{equation}
j_3^n=0, \,  \,
%j_4^n = \frac{N \left(\gamma _y-\gamma_c\right)}{4 \sqrt{\left( \gamma _x-\gamma_c\right) \left( \gamma _y-\gamma_c\right)}},
j_4^n = \frac{N}{4} \sqrt{\frac{\gamma _y-\gamma_c} {\gamma _x-\gamma_c}},
\end{equation}
and in the deformed phase, region II,
 \begin{eqnarray}
(j_3^d)^2=  -\frac{N^3 \gamma_c ^2 }{4 \gamma _x^3}\sqrt{\frac{\left(\gamma _x^2-\gamma_c ^2\right) \left(\gamma _x-\gamma _y\right)}{\gamma _x}} , \\ 
j_4^d = -\frac{N \left(\gamma _c{}^2 \left(5\gamma _x-3 \gamma _y\right)-\gamma _c \gamma _x \left(\gamma _y+3 \gamma _x\right)+2 \gamma _x^2 \gamma _y\right)} 
{8 \gamma _x \sqrt{\gamma _x \left(\gamma _x^2-\gamma_c ^2\right) \left(\gamma _x-\gamma _y\right)}}.
\end{eqnarray}

 Substituting in Eq. (\ref{chif}) we obtain
\begin{eqnarray}
\chi _{F}^n=&\frac{1}{32 \left(\gamma _x - \gamma_c\right)^2}, \\
\chi _{F}^d= &-\frac{ N }{ \sqrt{\left(\gamma _x - \gamma_c\right)}}
\frac{\gamma_c ^2}{4 \gamma _x^3 }
\sqrt{\frac{\gamma _x}{\left(\gamma_c +\gamma _x\right) \left(\gamma _x-\gamma _y\right)}} \nonumber\\
&+\frac{1}{\left(\gamma _c-\gamma _x\right)^2}
\frac{\left[\gamma _c \gamma _x \left(-5 \gamma _c+3 \gamma _x\right)
+\left(3 \gamma _c-2 \gamma _x\right) \left(\gamma _c+\gamma _x\right) \gamma _y\right]^2}{128 \gamma _x^2 \left(\gamma _c+\gamma _x\right)
{}^2\left(\gamma _x-\gamma _y\right)^2}.
\end{eqnarray}
Note that in the deformed region II $\gamma_x<\gamma_c=-\epsilon<0$ and $\gamma_x \le \gamma_y$, making the argument in the square roots, $(j_3^d)^2$ and $\chi _{F}^d$ positive. In the deformed region the first term is of order $N^1$ and the second of order $N^0$. For any value of $\gamma_x \ne \gamma_c$ the first term is the one to be considered in the thermodynamic limit  \cite{gu10,you07}.

The fidelity susceptibility $\chi _{F}$ exhibits very interesting features. It is divergent in both phases as $\gamma_x\rightarrow \gamma_c$, but with different powers of $(\gamma _x - \gamma_c)$ and of N in each phase \cite{gu10,you07}. Renormalising the fidelity susceptibility the following exponents are found
\begin{center}
\begin{tabular}{c|c|c|c}
& $\alpha$ &$\beta$ & $\beta + \alpha \, \nu$ \\ \hline
$\gamma_x \ge \gamma_c$ & 2 & 0 & 2$\nu$\\
$\gamma_x < \gamma_c $& $\frac 1 2$ & 1 & $1 + \frac{\nu}{2}$
\end{tabular}
\end{center}

The Lipkin model shows distinct critical exponents for the fidelity susceptibility around the critical point. The exact numerical calculations fully confirm that the critical exponents are different on both sides of the critical point. It can be clearly seen in the plots of the rescaled fidelity susceptibility against $N (\gamma_x - \gamma_c)$, which for any values of N falls exactly on the same line, with a noticeably asymmetry on both sides of the critical point \cite{gu10,you07,Cas_jerry}.

\begin{figure}[h]
\begin{center}
\scalebox{0.7}{\includegraphics{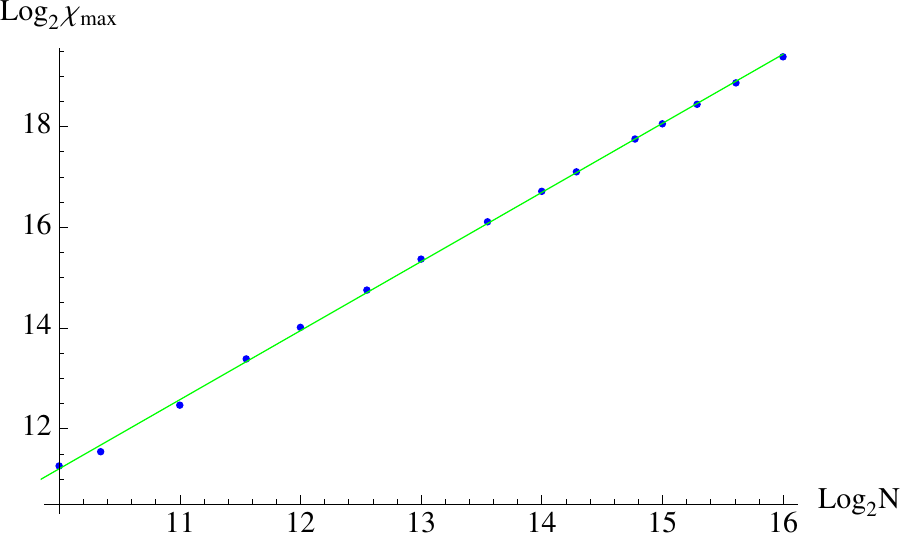}}
\scalebox{0.7}{\includegraphics{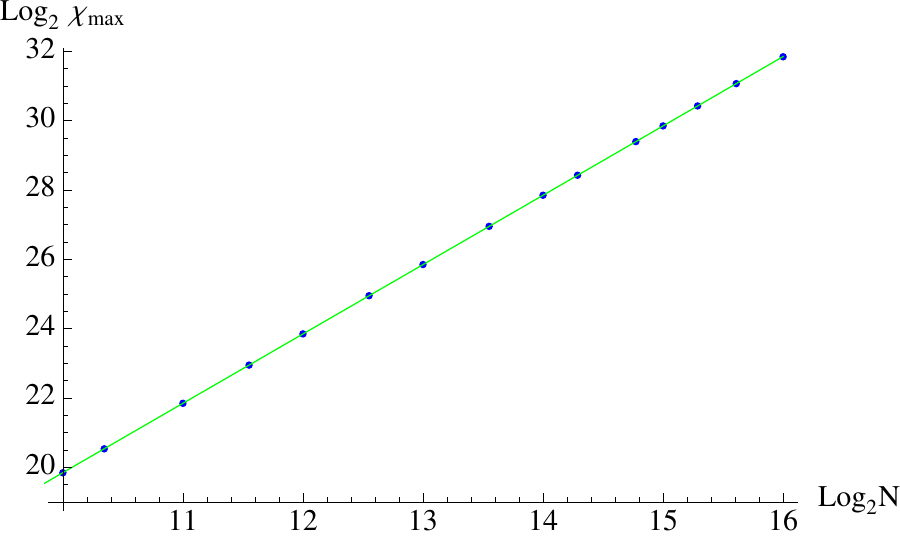}}
\end{center}
\caption{\label{fchi}
(Color online) Log-log plots of the maximum of the fidelity susceptibility as function of the number of particles N in the system, for $\gamma_c =-1$ and $\gamma_y= 1.0$ (left) and  $\gamma_y= -1.0$ (right). }
\end{figure}

It implies that the renormalized fidelity susceptibility $\chi_F$ scales as $N^{2\nu}$ in the normal region, and as $N^{1+ \frac{\nu}{2}}$ in the deformed region. As $\chi_F$ is continuos at the phase transition for any finite N, the two exponents should be equal, implying that
\centerline{$\nu = \frac 2 3$.}
In this way we have shown analytically that the Lipkin Hamiltonian belongs to this class of universality. It follows that $\chi_F$ diverges as $N^{\frac 4 3}$ at the phase transition. This dependence of $\chi_F$ at $\gamma_c$ is confirmed numerically \cite{Cas_jerry}. It is shown in the left of Fig. \ref{fchi}, where the maximum value of $\chi_F$ is plotted against $N$ in a log$_2$-log$_2$ plot.
The points are clearly along  a straight line, with slope $1.37 \approx 4/3$.  The numerical fit is $\chi_F= 0.175\, N^{1.37}$. The values of $N$ range from $2^{10}= 1024$ to $2^{16}= 65536.$ To the best of our knowledge, this is the first time that the discontinuity in the critical exponents of the fidelity susceptibility is employed to obtain analytically the class of universality associated with the Lipkin model.

The above deduction is valid for $\gamma_y > \gamma_c$ in the normal phase, region I, and for $\gamma_y < \gamma_x$ in the deformed phase, region II. There is a very singular behaviour along the line $\gamma_y = \gamma_c$ in the deformed region II.
In this case the fidelity susceptibility has the functional form 
\begin{equation}
\chi _{F}^d(\gamma_y = \gamma_c)= -\frac{ N }{ \left(\gamma _x-\gamma_c\right)}\frac{\gamma_c ^2}{4 \gamma _x^3 }\sqrt{\frac{\gamma _x}{\gamma _x+\gamma_c }}.
\end{equation}
Numerical studies \cite{Cas_jerry} have shown that $\nu(\gamma_y = \gamma_c)=1$, and $\chi_F\rightarrow N^2$ as it approaches the phase transition. This behaviour is plotted in the right of Fig  \ref{fchi}, with slope $2.001$.  The numerical fit is $\chi_F= 0.892\, N^{2}$ along the line $\gamma_y = \gamma_c, \gamma_x \rightarrow \gamma_{c\, -}$.

\section{Conclusions}

We have studied the Lipkin model of collective spins employing the Holstein-Primakoff mapping of the quantum rotor to a bosonic field. We have shown how the different phases of the system are found by using the coherent Heisenberg-Weyl state, which is an eigenstate of this boson. 
The possibility of going beyond the mean field description was reviewed. It was obtained by introducing a second set of displaced bosonic operators whose vacuum is the coherent state, expanding the square roots appearing in the mapping of the angular momentum operators as an infinite series, replacing it in the Hamiltonian, and truncating by keeping terms of order $N^1$, $N^{1/2}$ and $N^0$.  In this way a truncated Hamiltonian is built, which is quadratic in the new bosons, and can be diagonalized exactly by means of a Bogoliubov transformation. It provides a harmonic description of the Hamiltonian, with an excitation energy, the gap, which vanishes at the phase transition. It was shown that this description is consistent with the exact behaviour of the system at finite number of spins N, obtained through a numerical diagonalization of the full Hamiltonian.

On the other hand, it was also exhibited that the predictions for the ground state energy per particle and for the fraction of excited atoms exhibit a spike around the phase transition which is not present in the exact numerical calculations for a finite number of atoms.
%is spurious, an artifact of the truncation, which is itself not longer valid in a vicinity of the phase transition. 
The size of this spike goes, in the case of the energy per particle, to zero in the thermodynamic limit, while for the fraction of excited atoms is infinite, and remains infinite even in the limit $N \rightarrow \infty$. These singularities are a remainder that results obtained employing the truncated Hamiltonian for a finite of atoms
%, which inhibit the use of the truncated Hamiltonian 
close to the phase transition do not correspond to the exact ones, but carry anyway useful information. Through renormalisation, it was obtained  a recipe to build smooth and well behaved functions from singular ones by multiplying them with an appropriate function of the control parameter and the number of particles.

The behaviour for large N of the singular term in the energy per particle and of the number of excited atoms was obtained in this way. It was also shown that the fidelity susceptibility, an observable widely used in quantum optics, provides not only a very efficient tool to detect the precursor of the phase transition at finite N, but allows one to obtain analytically the class of universality to which the Lipkin model belongs: the one which is renormalized with a general factor $N^{2/3} (\gamma - \gamma_c)^{\alpha}$. The exponent $2/3$ was obtained asking for the fidelity susceptibility to be continuos at the phase transition,  and taking advantage of the fact that it has different functional forms in the different phases of the system.
It was also shown that there is a specific region, the line where $\gamma_y = \gamma_c$ and $\gamma_ x < \gamma_c$, in which the renormalizing function is different, i.e., the system behaves in a different way along this line. Numerical calculations confirm the N dependence of the fidelity susceptibility in both regions. 

We hope that this review of the subject would serve both as an introduction to the powerful techniques and to the care that must be taken in its use.

We thank J. Vidal for his comments. J.G.H. and E.N-A. thank P. Domokos and D. Nagy for their hospitality and valuable conversations. This work was partially supported by CONACyT-M\'exico and DGAPA-UNAM project IN102811.

\end{document}